\documentclass[epj]{svjour}

\newcommand{\bea}{\begin{eqnarray}}
\newcommand{\eea}{\end{eqnarray}}

\newcommand{\vect}[1]{\mathbf{#1}}

\newcommand{\di}{\displaystyle}

\newcommand{\ro}{r_{0,{\rm ref}}}
\newcommand{\zo}{z_{0,{\rm ref}}}
\newcommand{\agt}{\gtrsim}
\newcommand{\alt}{\lesssim}
\newcommand{\kbt}{k_{\rm B}T}
\newcommand{\mc}{\mathcal}

\usepackage{latexsym}
\usepackage{graphics,color}
\usepackage{epsfig}
\usepackage{amssymb}
\begin{document}
\title{Ellipsoidal particles at fluid interfaces}
\author{H. Lehle\inst{1}, E. Noruzifar\inst{2} \and M. Oettel\inst{2}% etc
}                     % Do not remove
\institute{Max-Planck-Institut f\"ur Metallforschung, Heisenbergstr. 3, 
  D-70569 Stuttgart and 
 Institut f\"ur Theoretische und Angewandte Physik, Universit\"at Stuttgart,
             Pfaffenwaldring 57, D-70569 Stuttgart, Germany
  \and Institut f\"ur Physik, WA 331, Johannes-Gutenberg-Universit\"at
  Mainz,
D-55099 Mainz, Germany}
\date{Received: / Revised version:}
\abstract{
 For partially wetting, ellipsoidal colloids trapped at a fluid interface, 
 their effective, interface--mediated interactions
 of capillary and fluctuation--induced type are analyzed.
 For contact angles different from 90$^o$, 
 static interface deformations arise which lead to anisotropic capillary 
 forces that are substantial already for micrometer--sized particles.
 The capillary problem is solved using an efficient perturbative treatment
 which allows a fast determination of the capillary interaction for all
 distances between and orientations of two particles.
 Besides static capillary forces, fluctuation--induced forces caused by 
 thermally excited capillary waves arise at fluid interfaces. For the specific choice
 of a spatially fixed three--phase contact line, the asymptotic behavior
 of the fluctuation--induced force is determined analytically for both the
 close--distance and the long--distance regime and compared to numerical solutions.  
 % end of PACS codes
} %end of abstract
\maketitle

\section{Introduction}

Colloidal particles trapped at fluid interfaces
exhibit effective interactions which broadly can be classified into ``direct"
interactions (e.g., of electrostatic or van--der--Waals type) which also
exist for colloids in bulk solvents but they are modified at interfaces. Additionally,
the presence of an interface gives rise to interactions mediated by deformations
of that interface, thus these are absent in bulk solutions of colloids. 
One speaks of capillary interactions \cite{Kra00}, if these deformations
are static, and of fluctuation--induced interactions if the deformations are
caused by thermal fluctuations. For spherical colloids at free interfaces, the
formation of self--organized structures is mainly governed by the direct interactions
whereas for particles of nonspherical shape capillary interactions appear to be
dominant \cite{Bre07r}.

Static interface deformations arise if the colloids experience forces in the 
direction parallel to the interface normal \textcolor{black}{(e.g. if they are 
pushed into the lower phase)} and/or stress
distributions act on the interface~\cite{Oet08}. 
For microcolloids of sizes less than 10 $\mu$m,
the omnipresent gravitational force on the colloids can be neglected.
\textcolor{black}{
However,
meniscus deformations also arise
in conjunction with direct interactions (like electrostatic forces)
which lead to 
forces and stresses 
on colloids and interface, respectively.} 
%can arise in conjunction with the direct interactions. 
If the system ``colloids + interface" is mechanically isolated (usually this
applies for colloid experiments in a Langmuir trough or on large droplets), then
the force on the colloids directed vertical to the interface is balanced by the
total force on the interface (obtained by integrating the stress distribution over
the interface area) \cite{For04,Dom05,Dom07}. 
The ensuing capillary interactions decay with
a power--law in the intercolloidal distance $d$, e.g., in the case of charged
colloidal spheres at air--water or oil--water interfaces they are attractive,
$\propto d^{-3}$, but for large $d$ they are usually weaker than the direct 
electrostatic repulsion which is also $\propto d^{-3}$ \cite{Oet05a,Wue05,Fry07}.  --
On the other hand, in the absence of forces on the colloids and stresses on the
interface,
static interface deformations can also be induced by an anisotropic colloid shape;
more precisely if the colloid is not symmetric with respect to rotations around
any axis through the colloid which is parallel to the normal on the undisturbed
interface. 
Young's equation requires that at the three--phase contact line the angle between
the local interface normal and the local normal on the colloid surface is given by the
contact angle $\theta$. Thus for an anisotropic colloid this condition 
cannot
 be
met if the  interface remains flat; the contact line will not be located in the
plane of the
undisturbed interface. The associated interface deformations around one such
colloid can be calculated in terms of a two--dimensional multipolar expansion
\cite{Fou02}. For asymptotically large distances from the colloid, the leading
nonvanishing multipole is in general the quadrupole, since monopole and dipole
are absent through the conditions of force and torque balance. The interaction
energy between two quadrupoles depends on the colloid orientation in the
interface plane and decays according to a power law, $\propto d^{-4}$. 
Experimentally,
it has become possible to produce anisotropic microcolloids of controlled shape
for investigations of their self--assembly at fluid interfaces: 
anorganic nanorods \cite{Yan02},
bended disks \cite{Ren01} or rotationally symmetric ellipsoids (spheroids) with
aspect ratios up to 10 \cite{Lou05,Bas06}. For the ellipsoids, results on the
effective
pair potential for intermediate distances $d$ \cite{Lou05} indicate a strong 
orientation dependence       
not captured by the aformentioned leading quadrupole interaction. Ellipsometric
measurements of the interface deformation around one particle are consistent
with a quadrupolar pattern \cite{Lou06} which, however, appears to be deformed
considerably for stretched ellipsoids.  

The calculation of the full interface profile around trap\-ped ellipsoids
and the evaluation of the associated capillary interaction energy can only be done
numerically, save for small eccentricities $e$ where an expansion in terms
of $e$ is possible \cite{Nie05}. Similarly, an evaluation of the interface profile is
simplified if for colloidal spheres only small, roughness--induced deviations
from the circular shape of 
the contact line are assumed \cite{Sta00,Dan05}. 
For ellipsoids with eccentricities not close to zero, however, the position of 
the contact line is not given {\em a priori}
but has to be determined self--consistently through Young's equation. This 
task is a variant of a free boundary problem which in general poses problems
in terms of speed and efficiency of a numerical algorithm aimed at solving it.  
Below, we show that a perturbative solution can be attained through expansion
of an appropriate free energy functional  and subsequent minimization 
which leads to a problem with fixed boundaries (Subsec.~\ref{sec:capsingle}).
The calculation of the capillary interaction between two ellipsoids
is thus greatly simplified and results can be obtained for the full range
of distances and orientations in a configuration
with two ellipsoids (Subsec.~\ref{sec:cappair}).      

The capillary interactions between two ellipsoids are absent if the contact
angle 
{$\theta$ equals $\pi/2$, and} 
they become significantly smaller for ellipsoid sizes
approaching the molecular length scale. In these circumstances, fluctuations 
around the equilibrium interface position are expected to influence the
effective interactions noticeably. Since the interface fluctuations
are of thermal nature, the scale of the ensuing interactions will be given
by $\kbt$, the thermal energy.        
Within a coarse--grained picture, the properties of fluid interfaces are
very well described by an effective capillary wave Hamiltonian
which governs both the equilibrium interface configuration
and the thermal fluctuations (capillary waves) around this equilibrium
(or mean-field) position.
As postulated by the
Goldstone theorem the capillary waves are long-range correlated.
The interface breaks the continuous translational symmetry
of the system, and in the limit of vanishing external fields -- like gravity --
it has to be accompanied by
easily excitable long wavelength (Goldstone) modes -- precisely
the capillary waves.
The fluctuation spectrum of the capillary waves will be modified by colloids
trapped at the interface and therefore leads to fluctuation--induced forces
between them. In that respect, colloids at fluids interfaces appear
to be a possible realization of a two--dimensional system exhibiting the Casimir
effect. 
For spheres and disks, these forces have been calculated 
in Refs.~\cite{Leh06,Leh07b}. The large--distance behavior of these
forces depends sensitively on the boundary conditions at the three--phase
contact line whereas for close distances a strong attraction similar
to van--der--Waals forces has been found, independent of the type of
boundary condition. 
In the case of colloidal rods the asymptotic behavior of the 
fluctuation--induced force
has been evaluated in Ref.~\cite{Gol96} and shown to lead to an orientational 
dependence.
Furthermore we note that there is numerous work on the force between inclusions
on membranes where the membrane shape fluctuations take the role of capillary 
waves, see, e.g., Refs.~\cite{Gol96,Goul93}.

In Sec.~\ref{sec:fluc} below, we consider the specific case of two ellipsoids
trapped at an interface with a pinned contact line. The fluctuation--induced 
force between two such ellipsoids corresponds to the Casimir force with Dirichlet
boundary conditions. We present numerical results for the whole distance regime
for selected orientations and compare to the leading terms in analytic expansions
valid for the close--distance and the long--distance regime.

Finally, Sec.~\ref{sec:sum} contains a discussion of the results.

\section{Static interface deformations and capillary interactions}

We consider cigar--shaped ellipsoids of micrometer size with half--axes $(a,b,b)$ 
and $a>b$ (\mbox{$e=(1-b^2/a^2)^{1/2}$} is the eccentricity)
trapped at an air--water interface with surface tension $\gamma$. 
The surface tensions of the ellipsoid with air and water are denoted by
$\gamma_{\rm I}$ and $\gamma_{\rm II}$, respectively. The contact angle (Young's
angle)
is defined by $\cos\theta = (\gamma_{\rm I}-\gamma_{\rm II})/\gamma$.
The free energy of the system shall be given only by the surface free energies
of the three involved interfaces (ellipsoid--air, ellipsoid--water and air--water),
i.e., we neglect gravitational effects which are negligible for micrometer--sized
particles and also possible electrostatic effects which arise through the
ubiquituous surface charges on real colloids. (We will comment upon electrostatic
effects in Subsec.~\ref{sec:cappair}).  

\subsection{Capillary deformation around a single ellipsoid}

\label{sec:capsingle}

\begin{figure}[t]
 \begin{center}
   \epsfig{file=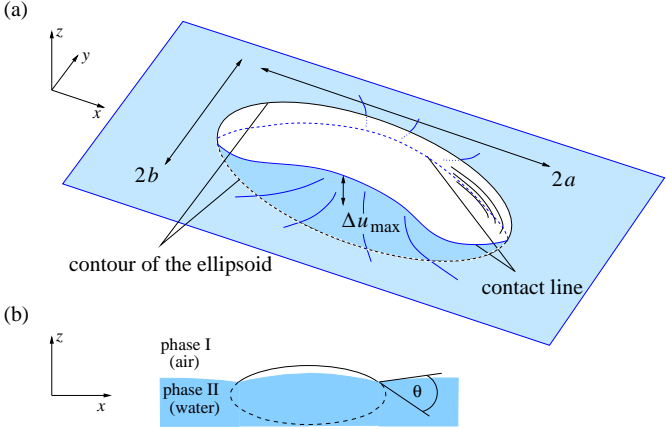, width=\columnwidth}
 \end{center}
 \caption{(a) Deformation of a fluid interface around an ellipsoid (satisfying Young's
equation
  locally), with $\Delta u_{\rm max}$ denoting
  the maximal difference in meniscus elevation along the three--phase
  contact line. The part of the ellipsoid protruding out of the water is shown
unshaded and
  the part of the ellipsoid immersed in water is depicted in darker shade.
\textcolor{black}{ (b) Side view of the ellipsoid with the geometrical definition of the contact 
     angle $\theta$.}  }
 \label{fig:ell1}
\end{figure}

For $\theta$ in the range between 0 and 180$^o$ and in the absence of line
tensions, the most stable configuration is given by the ellipsoid positioned flat on
the interface (Fig.~\ref{fig:ell1}), because in this configuration the amount of
displaced
area of the air--water interface is maximal.\footnote{This can be shown by using
the simplifying assumption that the interface around the ellipsoid remains flat 
\cite{Far03}.}
The meniscus deformation with respect to the plane $z=0$ is denoted by $u(\vect r)$
where $\vect r$ is a two--dimensional vector in the plane $z=0$   and
the vertical position of the ellipsoid center is given by $h$. In order
to determine the equilibrium deformation and vertical ellipsoid position, we
seek an expansion of the free energy around a reference configuration characterized
by the deformation $u_{\rm ref}(\vect r)$ and the vertical position $h_{\rm ref}$
such that
\bea
    u = u_{\rm ref} + v\;, \qquad h = h_{\rm ref} + \Delta h\;. 
\eea 
The properties of the reference configuration will be determined below. The change
in free energy with respect to this reference configuration is given by 
\bea
  \Delta {\cal F} = \gamma \Delta A_{\rm men} + \gamma_{\rm I} \Delta A_{\rm I}
      + \gamma_{\rm II} \Delta A_{\rm II}\;,
\eea    
where $\Delta A_{\rm men}$ denotes the change in meniscus area and 
$\Delta A_{\rm I[II]}$ denotes the change in contact area between the ellipsoid and
air or water, respectively. We split this free energy difference into two parts:
\bea
  \Delta {\cal F} &=&  {\cal F}_{\rm men} + {\cal F}_{\rm b}\;,\\
  \label{eq:fmen}
  {\cal F}_{\rm men} &=& \gamma \int_{S_{\rm ref}} d^2\vect r \left(
    \sqrt{1+(\nabla u)^2}-\sqrt{1+(\nabla u_{\rm ref})^2}\right) \;, \quad
\eea   
where the meniscus free energy ${\cal F}_{\rm men}$ denotes the difference of the
air--water interfacial energy integrated over that part of the
plane $z=0$ which is given by the area of the meniscus $u_{\rm ref}$ of the
{\em reference} configuration projected onto it (denoted by $S_{\rm ref}$).
The boundary free energy ${\cal F}_{\rm b}$ includes all remaining terms.
We denote by $S$ the area of the meniscus $u$ projected  onto the plane $z=0$ 
and $z_{\rm ell}(x,y;h)$ describes the surface equation of the ellipsoid which depends
on its vertical position $h$ as a parameter. With these definitions
\bea
 \label{eq:fbdef}
 {\cal F}_{\rm b} &=& \gamma \int_{\textcolor{black}{S_{\rm ref}\backslash S}}
d^2\vect r
  \left( \cos\theta \sqrt{1+(\nabla z_{\rm ell})^2}- \right. \\
  && \left. \qquad\qquad\qquad \textcolor{black}{\sqrt{1+(\nabla [u_{\rm ref}+v])^2}}
\right). \nonumber  
\eea
\textcolor{black}{At this point we perform a Taylor expansion up to second order 
in the meniscus deformation
for the free energy contribution  ${\cal F}_{\rm men}$ and ${\cal F}_{\rm b}$
separately. For ${\cal F}_{\rm men}$, this is equivalent to the small--gradient
expansion $|\nabla v|, |\nabla u_{\rm ref}| \ll 1$ and results in
\bea
 \label{eq:fmenapprox}
  {\cal F}_{\rm men}  \approx \frac{\gamma}{2} \int_{S_{\rm ref}} d^2\vect r
  \left( (\nabla[u_{\rm ref}+v])^2 - (\nabla u_{\rm ref})^2   \right)\;.
\eea 
The expansion of the boundary part ${\cal F}_{\rm b}$ is somewhat more involved.
Since the area of the domain $S_{\rm ref}\backslash S$ will be of second
order in the meniscus deformation, we can approximate  
$\sqrt{1+(\nabla [u_{\rm ref}+v])^2} \approx 1$ in Eq.~(\ref{eq:fbdef}) and find
}
\bea\label{Fbsingle}
 \label{eq:fbapprox1}
 {\cal F}_{\rm b} &  \approx & \gamma \int_0^{2\pi} d\phi \int_{\ro}^{ r(\phi)} 
   \!\!\!\!\!\!\!\!\!\!
 dr\, r
  \left( \cos\theta \sqrt{1+(\nabla z_{\rm ell})^2}- 1 \right) \;.
\eea
Here the functions $\ro(\phi)$ and $r(\phi)=\ro+\Delta r(\phi)$ 
parametrize
the polar radius of the boundaries $\partial S_{\rm ref}$ and
$\partial S$, respectively, i.e., they correspond to the projected 
three--phase contact lines formed by the reference meniscus $u_{\rm ref}$ and
the arbitrary meniscus $u$.
\textcolor{black}{
Subsequently we perform a functional expansion of Eq.~(\ref{eq:fbapprox1})} with
respect to 
displacements of the contact line position $\tilde v_\phi=v(r(\phi))-\Delta h$ which 
are constrained to lie on the ellipsoid surface:
\bea
 \label{eq:taylorfb}
 {\cal F}_{\rm b}[\tilde v] & \approx & 
  \int d\phi \left.\frac{\delta {\cal F}_{\rm b}}
  {\delta\tilde v_\phi}\right|_{\tilde v =0} \tilde v_\phi
  + \\
  & & \frac{1}{2} \int d\phi \int d\phi' \left. \frac{\delta^2 {\cal F}_{\rm b}}
  {\delta \tilde v_\phi \delta \tilde v_{\phi'}}\right|_{\tilde v =0} 
   \tilde v_\phi\tilde v_{\phi'}  + \dots \;.  \nonumber
\eea
The position of the reference contact line is fixed by the requirement that the
reference configuration minimizes ${\cal F}_{\rm b}$, i.e.,
\bea
 \left.\frac{\delta {\cal F}_{\rm b}}
  {\delta\tilde v_\phi}\right|_{\tilde v =0} =
  \left.\frac{d r(\phi)}{d\tilde v_\phi}\,\frac{\delta {\cal F}_{\rm b}}
  {\delta r(\phi)}\right|_{r(\phi) =\ro(\phi)}
  \stackrel{!}{=} 0 \;,
\eea 
which leads to the following condition on $\ro(\phi)$:
\bea
 \label{eq:r0}
  \cos \theta = \frac{1}{\sqrt{1+[\nabla z_{\rm ell}(\ro(\phi))]^2}} \;.
\eea
Geometrically, Eq.~(\ref{eq:r0}) expresses the condition that at the reference
contact line
the angle between the unit vector in $z$--direction and the ellipsoid normal
is given by $\theta$ which is a reasonable first ``guess" of the equilibrium
contact line position. The solution to this equation yields ellipses for
the projection of the reference contact line with half axes $b'=b|\sin\theta|$ and
$a'=b'/(1-e^2[1+\cos^2\theta(1-e^2)])^{1/2}$.
In order to fully specify the reference configuration, we require that
the reference meniscus is a surface with minimal area, i.e,
to first order it fulfills $\triangle u_{\rm ref}=0$ 
\textcolor{black}{(consistent with Eq.~(\ref{eq:fmenapprox}))} with the pinning
condition
$u_{\rm ref}|_{\ro(\phi)}= z_{\rm ell}(\ro(\phi);h_{\rm ref})$.

 The second term on the r.h.s. of
Eq.~(\ref{eq:taylorfb}) gives us the approximation for the boundary free energy
used in the following:
\bea
 \label{eq:fbapprox}
 {\cal F}_{\rm b}[\tilde v] & \approx & \frac{\gamma}{2} \int_0^{2\pi} d\phi\,
    R(\phi)\,\tilde v_\phi^2  \;, \\
 \label{eq:R}
   R(\phi) &=& \frac{b^2\sin^2\theta}{\ro(\phi)^2(1-e^2\cos^2\phi)}\;.
\eea 
\textcolor{black}{Since it is a local functional on the projected reference contact
line, 
this boundary free energy can be viewed as the total free energy cost in shifting
the contact line with respect to the reference state.} 
Thus we see that the minimization of the free energy 
$\Delta {\cal F}={\cal F}_{\rm men}+{\cal F}_{\rm b}$, Taylor expanded to second order
with ${\cal F}_{\rm men}$ given by Eq.~(\ref{eq:fmenapprox}) and ${\cal F}_{\rm b}$ 
by Eq.~(\ref{eq:fbapprox}), with respect to $v$ leads to the linearized
Young--Laplace equation $\triangle v =0$
(which also implies $\triangle u =0$ through $\triangle u_{\rm ref}=0$) 
 with the local boundary condition
on the projected reference contact line $\ro(\phi)$ 
\bea
 \label{eq:bc}
  \frac{\partial(u_{\rm ref}+v)}{\partial n} = -
    \frac{d\phi}{d\ell}\,R(\phi)\,( v(\ro(\phi))-\Delta h)\;.
\eea  
Here, $\partial/\partial n$ denotes the outward normal derivative on 
$\partial S_{\rm ref}\equiv \ro(\phi)$ and $d\ell$ is the line differential on
$\partial S_{\rm ref}$. Through minimization of $\Delta {\cal F}$ with respect to
$\Delta h$, the vertical position of the colloid is fixed by the condition
\bea
 \label{eq:dh}
   \Delta h = \frac{\di \int_0^{2\pi} d\phi\,R(\phi) \,v(\ro(\phi))}
              {\textcolor{black}{\di \int_0^{2\pi} d\phi\,R(\phi)}}\;.
\eea 
This condition implies that the solution does not depend on the choice
of $h_{\rm ref}$, the vertical position of the ellipsoid in the reference 
configuration. An arbitrary shift in $h_{\rm ref}$ will be compensated by 
a corresponding negative shift in $\Delta h$, as can be shown
from Eqs.~(\ref{eq:bc}) and (\ref{eq:dh}).  

At this point we want to remark that the reduction of 
the original
capillary problem (with unknown contact line) to the solution of the Laplace
equation  with a local boundary condition at a fixed boundary (the projection of the
reference contact line) is a great simplification and speeds up numerical solutions
enormously. Furthermore we note that the technique of splitting the free energy into
a meniscus and a boundary part with subsequent Taylor expansion has already been
introduced in Ref.~\cite{Oet05} for the problem of capillary deformations around
colloidal spheres.

Since the boundary curve $\ro(\phi)$ is itself an ellipse, a numerical solution
for the meniscus deformation $u$ is most conveniently performed using
elliptic coordinates $s,t$ whose relation to cartesian coordinates is given by
$x=\alpha\, \cosh s\cos t$ and $y=\alpha\, \sinh s\sin t$. Isolines
of constant $s$ are ellipses, the condition that the boundary curve
$\ro(\phi)$ is such an isoline at $s=s_0$ leads to the relations
$\alpha = a'e'$ and $s_0= \mbox{acosh}\, (1/e')$ where $e'$ is the eccentricity of the
elliptic boundary curve. The Laplace equation in elliptic coordinates is given by
\bea
  \triangle u (s,t)& =& \frac{1}{H^2}\left( 
  \frac{\partial^2}{\partial s^2} +  \frac{\partial^2}{\partial t^2}\right)
  \,u(s,t) = 0 \;, \\
   H &=& \alpha\sqrt{\sinh^2 s+\sin^2 t} \;,
\eea 
and its solution is given by the expansion
\bea
 \label{eq:mpell}
   u(s,t) &=& A_0\, \frac{s}{s_0} + \\
  & & \sum_{m>0} e^{-m(s-s_0)}
   \left[ A_m\,\cos(mt) + B_m\, \sin (mt) \right]\;, \nonumber
\eea
where $A_m$ and $B_m$ denote elliptic multipole moments of order $m$.
Comparing to the general solution in polar coordinates,
\bea
 \label{eq:mppol}
  u(r,\phi) &=& A_0^{\rm p} \ln\frac{r}{r_0} + \\
  & & \sum_{m>0} \left(\frac{r_0}{r}\right)^m\,
   \left[ A^{\rm p}_m\,\cos(m\phi) + B^{\rm p}_m\, \sin (m\phi) \right]\;, \nonumber
\eea 
we note that an elliptic multipole of order $m$ is a superposition
of polar multipoles of order $n\ge m$. In terms of the expansion given
in Eq.~(\ref{eq:mpell}), the problem reduces to a set of coupled linear equations
for the multipole moments. It is usually sufficient to take into account
multipoles of order $m \le 50$ for a very precise solution. 

\begin{figure}[t]
 \begin{center}
   \vspace*{5mm}
   \epsfig{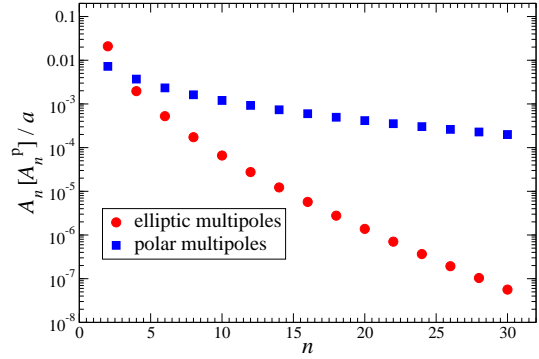}
 \end{center}
 \caption{Comparison of the elliptic and polar multipole expansion of
  the equilibrium meniscus around an ellipsoid with aspect ratio
  $a/b=5$ and contact angle $\theta=66^o$. The elliptic multipole series
  is defined in Eq.~(\ref{eq:mpell}) and the polar multipole series
  (where we have used $r_0=a$) is given by Eq.~(\ref{eq:mppol}). Note that
  because of symmetriy reasons all sine moments $B_m$ and $B_m^{\rm p}$ are zero.
 }
 \label{fig:multipoles}
\end{figure}

In Fig.~\ref{fig:multipoles} we compare the convergence of the elliptic and the
polar multipole series for the meniscus deformation around an ellipsoid
with aspect ratio $a/b=5$ and contact angle $\theta=66^o$.
The leading multipole is the quadrupole ($m=2$). The vanishing monopole
is related to the fact that no contact line force acts on the ellipsoid
which we have ensured by minimizing the free energy with respect to the
vertical position of the ellipsoid. The dipole moments are also zero since
in equilibrium there is no torque acting on the ellipsoid. 
 While the
elliptic multipole expansion converges sufficiently fast for all practical purposes
(e.g., $A_{30}/A_2 \approx 10^{-6}$), the polar multipole series is very badly
convergent (e.g., $A^{\rm p}_{30}/A^{\rm p}_2 \approx 0.03$). This is due to the
fairly large
aspect ratio; for $a/b \agt 1$ the polar multipole series yields rapid convergence
\cite{Nie05}.  

\begin{figure}[t]
 \begin{center}
   \vspace*{5mm}
   \epsfig{file=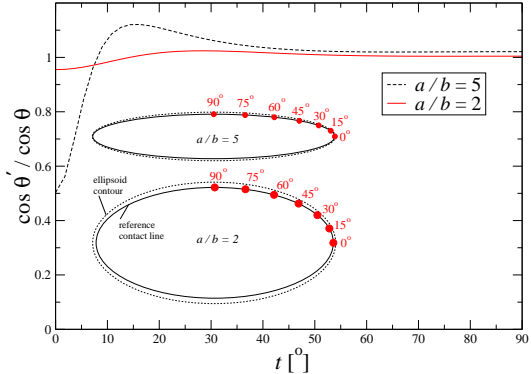,width=7cm}
 \end{center}
 \caption{\textcolor{black}{The ratio $\cos\theta'/\cos\theta$ for the two aspect ratios
2 and 5. The angle $\theta'$ is the contact angle pertaining to the
approximate solution (varying along the contact line contour, parametrized by the
elliptic
angle $t$) and $\theta= 66^o$ is the input contact angle.
The inset shows the projection of the ellipsoid contour and the reference contact line
on the interface plane. For some values of $t$, the location of the corresponding 
point on the contact line is given.
}}
 \label{fig:young}
\end{figure}

\textcolor{black}{
Due to the approximate forms of the meniscus and boundary free energy functionals
(Eqs.~(\ref{eq:fmenapprox}) and (\ref{eq:fbapprox})),
Young's condition is fulfilled only approximately for the corresponding minimum
configuration. In general, the approximation becomes exact in the limit 
$\theta \to 90^o$ (for arbitrary aspect ratio $a/b$) or $a/b \to 1$ (for arbitrary 
contact angle $\theta$). The projected contact line $r_0(\phi) \not = \ro(\phi)$ 
of the approximate solution  
can be determined numerically by the intersection of the solution (\ref{eq:mpell})
with the ellipsoid. The likewise numerically determined contact angle $\theta'$ varies
along $r_0(\phi)$, for an example see Fig.~\ref{fig:young} where 
the deviations from Young's law are shown for 
an input contact angle $\theta = 66^o$ and the  two aspect ratios $a/b=2$ and
$a/b=5$. For the smaller value of the aspect ratio, the deviations are
small. For $a/b=5$ however, larger deviations occur which are localized very close
to the tips (in a domain $t < 10^o$, see the inset of Fig.~\ref{fig:young}). Closer
inspection reveals that at the tips and on the contact line $r_0(\phi)$, 
the small--gradient
approximation $(\nabla u)^2 \ll 1$ breaks down (however, it still holds
approximately on the  {\em reference} contact line $\ro(\phi)$).
}

\begin{figure}[t]
 \begin{center}
   \vspace*{5mm}
   \epsfig{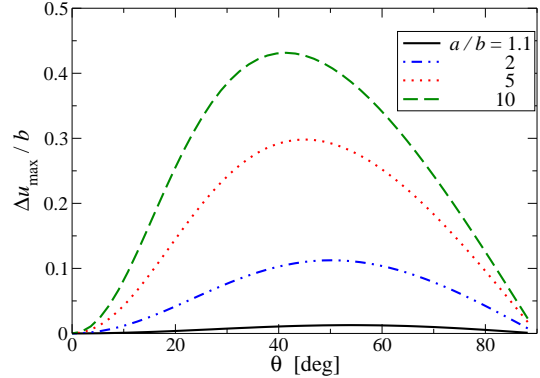}
 \end{center}
 \caption{Maximum meniscus height difference $\Delta u_{\rm max}(\theta)$
   as function of the contact angle $\theta$ for
   different aspect ratios $a/b$. 
   $\Delta u_{\rm max}$ attains its maximum for contact angles
   between 40$^o$ and 55$^o$, while it is zero for neutrally ($\theta=90^o$)
   or completely ($\theta \to 0$) wetting particles.
}
 \label{fig:du}
\end{figure}

The meniscus shape strongly depends on the aspect ratio and the contact angle.
We have investigated the influence of these parameters on the maximum height
difference 
$\Delta u_{\rm max}$ along the contact line, with the results shown in 
Fig.~\ref{fig:du}.  This provides us with a quick estimate on the strength
of the  interaction between two ellipsoids since in linearized theory one can
expect that the amplitude of the capillary interaction energy is 
approximately proportional to $(\Delta u_{\rm max})^2$. 
Depending on the precise value of the
aspect ratio, $\Delta u_{\rm max}$ attains its maximum for contact angles
between 40$^o$ and 55$^o$, i.e. in the experiments of Ref.~\cite{Lou06}, 
where for the used polystyrene ellipsoids a contact angle of around 40$^o$
was determined, the capillary deformation around the particles is close to its
maximum as compared to particles of the same shape but with different $\theta$.

\subsection{Capillary interaction between two ellipsoids}
\label{sec:cappair}

\begin{figure}[t]
 \begin{center}
   \epsfig{file=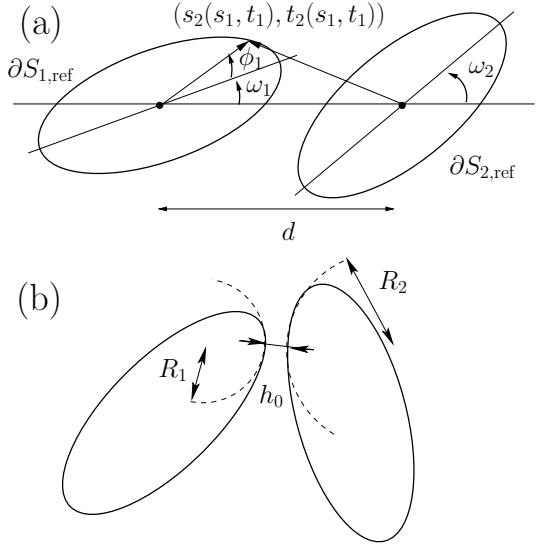,width=7cm}
 \end{center}
 \caption{
Top view on two ellipsoids at an interface. (a) The polar angles
$\omega_1$ and $\omega_2$ specify the direction of the long half axis of the contact
line ellipse
with respect to the  distance vector between the  centers of the two ellipses.
\textcolor{black}{
The polar angle $\phi_1$ on the reference ellipse $\partial S_{1,\rm ref}$
and the parametrisation $(s_1,t_1)=(s_1(s_2,t_2),t_1(s_2,t_2))$
of $\partial S_{1,\rm ref}$ in terms of elliptic coordinates
with respect to the reference ellipse $\partial S_{2,\rm ref}$ is indicated.
}
(b) $R_1$ and $R_2$ are the radii of the curvature on those ellipse points
whose distance is the minimal distance $h$. 
}
 \label{fig:ellfluc}
\end{figure}

For large distances $d$, the capillary interaction between two ellipsoids is 
determined by the quadrupole. For small eccentricities ($a \simeq b = r_0$)
the interaction energy has been determined in Refs.~\cite{Sta00,Fou02} and 
reads
\bea
 \label{eq:quad}
   U_{\rm cap}^{\rm quad} = -3\pi\gamma\, (\Delta u_{\rm max})^2\;
  \left(\frac{r_0}{d}\right)^4\;\cos(2\omega_1+2\omega_2)\;,
\eea
where $\omega_1$ and $\omega_2$ are the polar orientation angles of the ellipsoids
in the interface plane with respect to the distance vector between their centers
(see Fig.~\ref{fig:ellfluc} (a)). 
 The simple energy estimate in Eq.~(\ref{eq:quad})
predicts that the corresponding forces between two ellipsoids approaching each other
side--by--side
($\omega_1=\omega_2=90^o$)
or tip--to--tip ($\omega_1=\omega_2=0^o$) are attractive and equal. 
However, an experimental
estimate of these interaction forces
\cite{Lou05} revealed that the attractive force in the
side--by--side configuration varied as $F_{\rm cap} \propto -d^{-4.1}$ whereas
for the tip--to--tip configuration it varied as
$F_{\rm cap} \propto -d^{-5}$, in accordance with Eq.~(\ref{eq:quad}). The
measurements were performed over a limited distance range $d/a \alt 4$ for
aspect ratios $a/b$ ranging from 3 to 4.3.

Thus, for such aspect ratios the quadrupole approximation appears to be 
insufficient. 
We can apply the formalism developed in the previous subsection
also to the evaluation of the interface deformation and associated capillary energy 
for two ellipsoids.
 The meniscus part of the free energy (Eq.~(\ref{eq:fmen}))
remains unchanged save for the extend of the integration domain $S_{\rm ref}$:
here it consists of the whole plane with the {\em two} ellipses enclosed by the
projected reference contact line cut out,
\textcolor{black}{
$\hat S_{\rm ref} =\mathbb{R}^2
\setminus \cup_{i=1}^2 S_{i,\rm ref}$.\footnote{We follow the convention of 
Ref.~\cite{Oet05} and denote with a hat all quantities pertaining to the 
two--ellipsoid configuration.}}
 The boundary free energy for each ellipsoid
(Eq.~(\ref{eq:fbapprox})) has to be amended by additional terms related to
the appearance of an additional degree of freedom
given by the angles  $\alpha_i$ 
of the long ellipsoid axis with  the plane $z=0$. 
For fixed distance $d$ and orientation angles $\omega_1$ and $\omega_2$, 
the presence of the second
ellipsoid causes the first ellipsoid to dip with its tip into the water and
vice versa, i.e., 
the free energy has to be minimized with respect to rotations
around an axis in the interface plane which is perpendicular to the symmetry axis.
\textcolor{black}{
The additional terms in the boundary free energy can be determined as before
in Sec.~\ref{sec:capsingle}
by performing a Taylor expansion in terms of both, 
displacements of
the contact line
height $\tilde v_{i,\phi_i}=\hat v(r_0(\phi_i))-\Delta h_i$ and changes of the tilt
angle $\alpha_i$
% of the long ellipsoid axis
around the reference configuration $\tilde{v}_{i,\phi_i}=0$
and $\alpha_i=0$ 
(i.e., in the reference configuration the ellipsoids
are positioned flat on the undisturbed interface). 
This procedure %is performed in the appendix and
results in the expression
\bea\label{Fb2}
\hat \mc{F}_{\rm b}
&=&
\frac{\gamma}{2}
\sum_{i=1}^2\int_0^{2\pi} d\phi_i
\left\{R_{zz}(\phi_i)\,[\hat{v}-\Delta h_i]^2
\right.\nonumber\\&&\left.{}
+2 e^2 x R_{z\alpha}(\phi_i)  \,[\hat{v}-\Delta h_i]\,\alpha_i
+e^2 R_{\alpha\alpha}(\phi_i)\, \alpha_i^2
\right\} \quad 
\eea
for the total boundary free energy of  two ellipsoids.
The  lengthy expressions for 
the coefficients $R_{zz}$, $R_{z\alpha}$ and $R_{\alpha\alpha}$
are given in the appendix, see Eqs.~(\ref{Ellcoeffzz})--(\ref{Ellcoeffaa}).
(Note that $R_{zz}(\phi)=R(\phi)$, Eq.~(\ref{eq:R}).)
}

\textcolor{black}{For two ellipsoids, the meniscus free energy of the reference 
configuration, 
\bea
 \hat\mc{F}_{\rm men,ref} (d,\omega_1,\omega_2) \approx
  \frac{\gamma}{2} \int_{\hat S_{\rm ref}} \!\!\!d^2x\, 
  (\nabla \hat u_{\rm ref})^2 \;,
\eea
obviously depends on the configuration variables $\{d,\omega_1,\omega_2\}$.
The reference meniscus fulfills $\triangle \hat u_{\rm ref} =0$ and
the boundary condition
$\hat u_{\rm ref}|_{\partial S_{i,{\rm ref}}}=z_{\rm Ell}(r_{0}(\phi_i);h_{i,{\rm
ref}})$.
To include all dependence on the configuration variables into the total free 
energy, it is suitable to consider the free energy difference with respect to
the reference state with $d\to\infty$:
\bea
 \Delta \hat\mc{F} &=& \hat\mc{F}_{\rm men}(d,\omega_1,\omega_2) - 
   \hat\mc{F}_{\rm men,ref}(d \to \infty,\omega_1,\omega_2)
  + \nonumber \\
  & & \hat\mc{F}_{\rm b}(d,\omega_1,\omega_2) \\
 \label{FmenEll}
  &\approx & 
\frac{\gamma}{2}
\int_{S_{\rm ref}}
\!\!\!d^2x\,(\nabla \hat u)^2
+
\hat \mc{F}_{\rm b}(d,\omega_1,\omega_2) + {\rm const.}\;
\eea
Consequently, the capillary potential $U_{\rm cap}$ can be defined as the 
free energy difference
\bea
U_{\rm cap} 
=
\Delta\hat\mc{F}(d,\omega_1,\omega_2) - \Delta\hat\mc{F}(d \to
\infty,\omega_1,\omega_2)\;.
\eea
}

\textcolor{black}{
The equilibrium configuration minimizes the free energy
in Eq.~(\ref{FmenEll}) and can be calculated similarly as in 
Sec.~\ref{sec:capsingle}.
Minimizing $\hat\mc{F}_{\rm b}$ with respect to $\Delta h_i$
and $\alpha_i$ provides the equilibrium height
and orientation of ellipsoid $i$ and leads to 
 expressions analogous to Eq.~(\ref{eq:dh})
(since terms coupling $\Delta h_i$ and $\alpha_i$ vanish).
In contrast to the single colloid case,
the equilibrium meniscus $\hat u({\bf x})$ 
for two ellipsoids
has to fulfill
boundary conditions at {\it both} reference
ellipses $\partial S_{i,\rm ref}$.
%which
They
arise by minimizing the boundary free energy $\hat\mc{F}_{\rm b}$ in Eq.~(\ref{Fb2})
and contain
 an additional, $\alpha_i$-dependent term as compared to the
boundary condition for the single ellipsoid (Eq.~(\ref{eq:bc})).
For the numerical determination of the equilibrium meniscus 
profile, 
the superposition ansatz
$\hat{u}({\bf x}) =  u_1(s_1,t_1)+u_2(s_2,t_2)$ is used.
Thereby, the functions $u_i$
are given by the 
expansions of the meniscus around a single
colloid into elliptic multipoles (Eq.~(\ref{eq:mpell})),
and $(s_i,t_i)$ are elliptic coordinates with respect to 
the center of the reference
ellipse $S_{i,\rm ref}$ (see Fig.~\ref{fig:ellfluc}).
Through the boundary conditions %, which
%arise by minimizing the boundary free energy in Eq.~(\ref{Fb2}), 
a set of coupled linear equations
for the multipole moments is derived which can be solved with standard numerical
methods. 
}

\begin{figure}[t]
 \begin{center}
   \vspace*{5mm}
   \epsfig{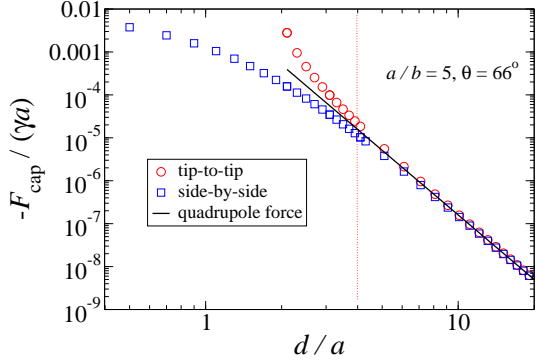}
 \end{center}
 \caption{
   Capillary force $F_{\rm cap}=-\partial U_{\rm cap}/\partial d$ in units of
$\gamma a$
   between two ellipsoids with aspect ratio $a/b=5$ and contact angle
   $\theta =66^o$
   approaching each other
   side--by--side or tip--to--tip. }
 \label{fig:ell2}
\end{figure}

As an example for the results, the capillary force 
$F_{\rm cap}=-\partial U_{\rm cap}/\partial d$
in direction of the distance vector between the ellipsoids is shown
in Fig.~\ref{fig:ell2}, for an aspect ratio $a/b=5$ and a contact angle 
$\theta=66^o$. 
The force considerably deviates from
the quadrupole form for $d/a <4$, i.e., in the region where the experimental
measurements of Ref.~\cite{Lou05} have been performed. 
For these distances, the force does not follow a power law but
a fit to an effective power--law would clearly yield an exponent $> -5$ 
in the side--by--side
configuration, but
also an effective exponent $<-5$ for the tip--to--tip configuration.
The reason for this behavior appears to be that the capillary deformation
around one ellipsoid is dominated by the  elliptic quadrupole, i.e.,
closer to the ellipsoid the capillary deformation
has substantial contributions from polar multipoles higher than the quadrupole
(see Fig.~\ref{fig:multipoles}) which also influence the pair interaction
considerably.

For the parameters used in Fig.~\ref{fig:ell2} the asymptotic capillary potential,
given by the quadrupole form $U_{\rm cap}^{\rm quad}=-U_0\,(a/d)^4\,
\cos(2\omega_1+2\omega_2)$, one finds the amplitude $U_0 \approx 7\times 10^6$
$\kbt$ for ellipsoids with long half axis $a=10$ $\mu$m at the air--water interface.
Usually, the experimentally used ellipsoids are charge--stabilized which leads
to an asymptotically isotropic dipolar repulsion $U_{\rm el} = U_{0,{\rm el}}\,
(a/d)^3$. Using the results of Ref.~\cite{Fry07}, one can estimate 
the amplitude of the electrostatic repulsions as $U_{0,{\rm el}} \approx 10^3\;\kbt$
(with a charge density of 1 electron per nm$^2$ and ultrapure water). Thus
the electrostatic repulsions are completely unimportant compared with the
capillary potential; only for distances $d \agt 10^4\, a$ the directional capillary
attractions would be overpowered by the electrostatic repulsions.

\section{Capillary waves: fluctuation--induced interactions}

\label{sec:fluc}

\begin{figure}[t]
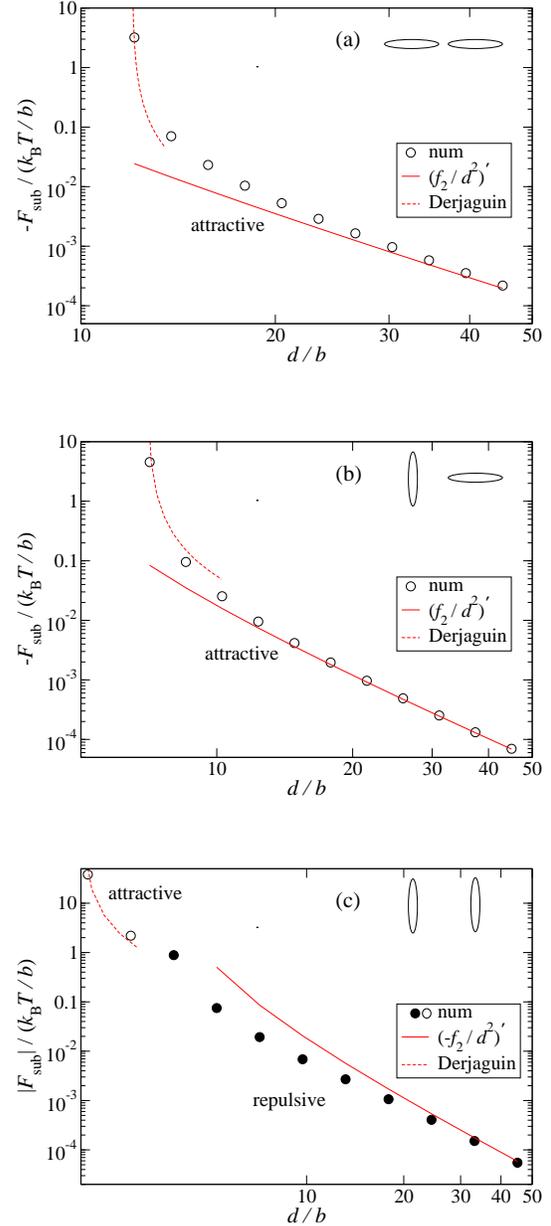

 \begin{center}
   \vspace*{5mm}
   \epsfig{file=600.eps, width=7cm} \\ \vspace*{9mm}
   \epsfig{file=6pi20.eps, width=7cm}\\  \vspace*{9mm}
   \epsfig{file=6pi2pi2.eps, width=7cm}
 \end{center}
 \caption{Results for the fluctuation force with the leading asymptotic term
subtracted,
$F_{\rm sub}=F_{\rm fluc} + \kbt\,\partial f_0/\partial d$, for ellipsoids
with aspect ratio $a/b=6$ and for the
three configurations (a) tip--to--tip ($\omega_1=\omega_2=0^o$), (b)
side--to--tip ($\omega_1=90^o$, $\omega_2=0^o$) and (c) side--by--side
($\omega_1=\omega_2=90^o$). Numerical results are shown by circles, the 
next--to--leading asymptotic term involving the coefficient $f_2$ (Eq.~(\ref{eq:f1}))
is represented by a full line, and the Derjaguin approximation (derivative of
Eq.~(\ref{eq:derjag}) with respect to $h_0$) is given by a dashed line, respectively.
The capillary length was chosen as $\lambda_c = 10^6\,b$.}
 \label{fig:ellfluc1}
\end{figure}

As discussed in the introductory section, the scale of the fluctuation--induced 
interaction energies is $\sim \kbt$ such that they are observable only if static
capillary interactions are (almost) absent. Since for $\theta=90^o$
the capillary interactions are identically zero (see the previous section) we discuss
in this section the exemplary case of two ellipsoids with their centers at 
$z=0$ and their
three--phase contact lines (ellipses with half axes $a$ and $b$) also located
in the plane of the flat interface, corresponding to the equilibrium
position of an ellipsoid with contact angle $\theta=90^o$.
Furthermore we assume that the contact line is pinned and the position of the 
ellipsoid is fixed by some external means. This corresponds to a Dirichlet boundary
condition for the meniscus at the boundary of the projected  meniscus area $S$:
$u(\partial S) =0$. The free energy cost for small--gradient fluctuations of the 
interface position $u(\vect r)$ around its mean $u=0$ is given by the capillary wave 
Hamiltonian:
\bea
 \label{eq:hcw}
  {\cal H}_{\rm cw} = \frac{\gamma}{2} \int_S d^2\vect r \left[ (\nabla u)^2
   +\frac{u^2}{\lambda_{c}^{2}} \right]\;,
\eea
which (with $\lambda_{c} \to \infty$) has already been used for evaluating 
the free energy of static interface deformations (see Eq.~(\ref{eq:fmenapprox})).
In Eq.~(\ref{eq:hcw}), $\lambda_{c}$ denotes the capillary length which is usually 
much larger than the extensions of microcolloids, nevertheless it is necessary to
keep the associated free energy contribution (stemming from gravity) throughout 
the calculations since it
ensures the stability of the interface \cite{Leh07b}. The capillary wave Hamiltonian
depends on the position and orientation of the ellipsoids through the integration
domain $S$ which encompasses the whole plane $z=0$ with the ellipses enclosed
by the contact lines cut out.
Therefore also the the free energy
${\mathcal F}(d,\omega_i)= -k_{\rm B} T \;\ln {\mathcal Z}(d,\omega_i)$
depends on the distance $d$ and the orientation angles $\omega_1$ and $\omega_2$
(see Fig.~\ref{fig:ellfluc} (a)).
The partition function
${\mathcal Z}(d,\omega_i)$ is obtained by a functional integral over all 
possible interface
configurations $u$; the boundary condition $u|_{\partial S_i}=0$ on the
two contact lines $\partial S_1$ and $\partial S_2$ is included by 
$\delta$-function constraints, as introduced in Ref.~\cite{Li91}:
\begin{equation}\label{Z1}
\mathcal{Z}=\mathcal{Z}_0^{-1} \int \mathcal{D}u\,
\exp \left\{-\frac{\mathcal{H}_{\rm cw}[u] }{k_{\rm B}T} \right\}\;
\prod_{\rm i=1}^2 %\int dh_i
\prod_{{\bf x}_i \in \partial S_{i}}
\delta [u({\bf x}_{ i})]
\;.
\end{equation}
$\mathcal{Z}_0$ is a normalization factor such that $\mathcal{Z}(d\to\infty)=1$.
The $\delta$-functions in Eq.~(\ref{Z1})
can be removed by using their integral representation via
 auxiliary fields $\psi_i ({\bf x}_i)$ defined on the
contact lines $\partial S_i$ ~\cite{Bor85,Li91}. This enables us to integrate out
the field $u$ leading to
\begin{eqnarray}
\label{aux}
 \mathcal{Z}& = & \mathcal{Z}_0^{-1} 
\int \prod_{i=1}^2 \mc{D}\psi_i\,
\exp\left\{
-\frac{k_{\rm B}T}{2\gamma}\sum_{i,j=1}^2
\oint_{\partial S_i}d\ell_i \oint_{\partial S_j}d\ell_j\, \times \right. \nonumber \\
 & & \left. \phantom{\frac{1}{2}} \qquad \qquad
\psi_i ({\bf x}_i)\,G(|{\bf x}_i-{\bf x}_j|)\,\psi_j({\bf x}_j)\right\}\;.
\end{eqnarray}
Here, we introduced the Green's function\\
$G({\bf x})=K_0(|{\bf x}|/\lambda_{  c})/(2\pi)$
of the operator $(-\triangle + \lambda_c^{-2})$
where $K_0$ is the modified Bessel function of the second kind.
In this form, the fluctuation part  resembles 2d screened electrostatics: it is the
partition function of a system
of fluctuating charge densities $\psi_i$ residing on the contact lines. 

In the intermediate asymptotic regime $a,b \ll d \ll \lambda_c$ the free energy
associated with this partition function can be calculated through an expansion
of the auxiliary fields $\psi_i$ in terms of elliptic multipoles
and an expansion of Green's function in terms of $1/d$ using elliptic coordinates.
Details of the lengthy calculations will be reported elsewhere.
The resulting free energy can be written as an expansion 
${\cal F}/(\kbt) = f_0 + f_2/d^2 + f_4/d^4 + \dots $ with the two leading coefficients
given by:
\bea
 \label{eq:f0}
   f_0  & = & \frac{1}{2} \ln \ln \frac{2d}{a+b} + {\rm const.} \;, \\
 \label{eq:f1}
   f_2  & = & -\frac{(a+b)^2}{4} - \frac{3}{8}(a^2-b^2)( \cos 2\omega_1+
              \cos 2\omega_2) \;.  
\eea
These expressions have been obtained in the limit $d/\lambda_c\to 0$ which, however,
is attained very slowly. In this limit the free energy difference
${\cal F}(d,\omega_i)-{\cal F}(d\to\infty,\omega_i)$ is actually
ill--defined 
and therefore the effective colloidal interaction 
is only meaningful for a finite capillary length
$\lambda_c$,
similar to  a free two--dimensional interface the width
of which is determined by the
capillary wave fluctuations and diverges logarithmically $\sim\ln \lambda_c$.
The leading terms of the fluctuation--induced
free energy between spheres or disks (where the contact lines are circles
of radius $r_0$) have already been calculated in Refs.~\cite{Leh06,Leh07b}
and correspond to the orientation--independent terms in Eqs.~(\ref{eq:f0}) and 
(\ref{eq:f1}) with $2r_0=a+b$. Note that for stretched ellipsoids with 
$a>2b$ the next--to--leading order term $f_2$ in the free energy expansion
may become repulsive when the ellipsoids are aligned side by side
($\omega_1=\omega_2=\pi/2$).

In the
opposite limit of small surface--to--surface distance $h_0(d,\omega_1,\omega_2)$ the
fluctuation force can be calculated by using the
well--known result for the fluctuation force per length $f_{\rm 2d}(\tilde h)=-
k_{\rm B}T\,\pi/(12 {\tilde h}^2)$
between two lines
a distance $\tilde h$ apart
\cite{Li91}, together with the Derjaguin (or proximity) approximation:
\bea
 \label{eq:derjag}
   \frac{{\cal F}}{\kbt} \approx  -\frac{\pi^2}{24} \sqrt{\frac{2}{h_0 
   \left( R_1^{-1} + R_2^{-1} \right)}} \qquad (h_0 \to 0)\;.
\eea
Here, $R_1$ and $R_2$ are the radii of curvature at those points on the contact line
of ellipsoid 1 and 2, respectively, whose distance is the minimal
surface--to--surface distance $h_0$ (see Fig.~\ref{fig:ellfluc} (b)). 
Thus, the fluctuation--induced interaction energy between the ellipsoids diverges
upon approach, similarly to the van--der--Waals attraction.
% which varies $\propto 1/h$ for $h \ll b$.  

For intermediate distances $d$ the partition function must be evaluated numerically.
In Eq.~(\ref{Z1}) the integral
over the auxiliary fields $\psi$ can be carried out because they appear only 
quadratically in the
exponent. The resulting determinant is divergent and requires regularisation. 
However, the
derivative of its logarithm with respect to $d$ (corresponding to minus the force
$F_{\rm fluc}$ in 
direction of the distance vector between the centers of the ellipsoids)
is finite and convergent
in a numerical analysis (see Ref.~\cite{Leh07b} for further details). 
It turns out that the fluctuation force is attractive for all distances and 
orientations which were analyzed. This is already suggested by the close--distance
regime (where $F_{\rm fluc} \propto -1/h_0^{3/2}$ is always attractive) and the
long--distance regime (where $F_{\rm fluc}$ is dominated by the likewise 
attractive, in--plane isotropic
term $-\partial f_0/\partial d = -1/[2d\ln(d/r_0)]$, see Eq.~(\ref{eq:f0})).
In order to exemplify the effect of in--plane anisotropy on the
fluctuation force, the results for the force with the asymptotically leading,
iso\-tropic
term subtracted ($F_{\rm sub}=F_{\rm fluc} + \kbt\,\partial f_0/\partial d$) 
are shown in Fig.~\ref{fig:ellfluc1} for ellipsoids
with aspect ratio $a/b=6$ and for the
three configurations (a) tip--to--tip ($\omega_1=\omega_2=0^o$), (b)
side--to--tip ($\omega_1=90^o$, $\omega_2=0^o$) and (c) side--by--side
($\omega_1=\omega_2=90^o$). For all configurations, for large $d$ the approach to the
aymptotic result given by $-\partial(f_2/d^2)/\partial d$ is fairly slow.
For the configurations (a) and (b) the subtracted force $ F_{\rm sub}$ 
remains attractive for all 
distances and there is a smooth crossover from the longe--distance
to the close--distance regime  while for the side--by--side configuration 
(c) there is a sign change
from the attractive close--distance regime (open circles) to the repulsive
long--distance regime (full circles), in accordance with Eq.~(\ref{eq:f1}).

\section{Summary and conclusions}

\label{sec:sum}

In this work we have analyzed the interface--mediated interactions which arise
between ellipsoidal particles trapped at a fluid interface. 

Firstly,
ellipsoids cause static interface deformations  if they are partially wetting and
their contact angle is different from 90$^o$. These static deformations lead 
to orientation--dependent capillary interactions between the particles.
The full solution to this capillary problem requires the solution of a nonlinear
differential equation together with Young's condition on the 
boundary, the three--phase contact lines whose locations are {\em a priori} unknown.
It is possible to analyze the interface deformation and the ensuing capillary
potential in a perturbative treatment, valid for small deformations of the
interface,  
which leads to the standard problem of a linear differential equation with
a local condition on a given, fixed boundary. For small to intermediate distances
between the ellipsoids we find considerable deviations from the well--known
quadrupole interaction which is valid for asymptotically large distances. 
As a perspective for future work,
the developed algorithm allows a fast determination
of the deformation and the potential also for large eccentricities of the particles
and appears to be potentially useful for application in computer simulations
of the aggregation process in ellipsoidal monolayers.  

Secondly, thermally excited capillary wave cause fluc\-tuation--induced interactions
between the ellipsoids. For the specific case of a pinned contact line we find
that anisotropic effects in the fluctuation force arise only for subleading terms
in an asymptotic expansion. It diverges for ellipsoids coming close to contact.
However, due to its small scale  the fluctuation force appears to be 
relevant experimentally only if the static capillary interactions are greatly reduced,
e.g., if the ellipsoids are of nanometer size or the contact angle is close to 
90$^o$.   

{\bf Acknowledgment:} E.~Noruzifar and M.~Oettel thank the German Science 
Foundation for financial
support \\ through the Collaborative Research Centre (SFB-TR6) 
``Colloids in External Fields".

\appendix

\section{Contact line contributions to the free energy}
\textcolor{black}{
In this appendix we determine the coefficients 
$R_{zz}$, $R_{z\alpha}$ and $R_{\alpha\alpha}$
of the functional Taylor expansion of the boundary free
energy $\hat\mc{F}_{\rm b}$ in Eq.~(\ref{Fb2})
around the reference configuration 
$\{\tilde{v}_{i,\phi}=\hat v(r_0(\phi_i))-\Delta
h_i=0,\alpha_{i,\rm ref}=0\}$.
They are given by the second variation of $\hat\mc{F}_{\rm
b}=\sum_{i=1}^2\mc{F}_{{\rm b},i}$
with respect to shifts of the contact line height $\tilde{v}_{i,\phi_i}$
or to changes in the orientation $\alpha_i$ of the long axis
of ellipsoid $i$ and can be calculated separately for the two colloids.
As in Sec.~\ref{sec:capsingle}
for the case of $\alpha_i\equiv 0$ (for a single ellipsoid),
the needed
 functional derivative 
$\delta^2 \mc{F}_{{\rm b},i}/\delta {\tilde{v}_{i,\phi_i}}^2$,
the derivative 
$\partial^2 \mc{F}_{{\rm b},i}/\partial {\alpha_i}^2$
and the mixed derivative
$ \frac{\partial}{\partial \alpha_i}
(\delta \mc{F}_{{\rm b},i}/\delta {\tilde{v}_{i,\phi_i}}$)
are 
determined by the derivatives of 
the boundaries
of the surface integrals 
(given by
the position of the three phase contact line,
cf. Eq.~(\ref{Fbsingle}))
with respect to $\alpha_i$ and $\tilde{v}_{i,\phi}$, respectively.
}

\textcolor{black}{
According to Eq.~(\ref{eq:fbapprox1}),
$\mc{F}_{{\rm b},i}=\gamma(\cos\theta \Delta A_{{\rm I},i} +\Delta
A_{{\rm
  proj}, i})$,
upon contact line shift and tilt the boundary free energy 
contains a contribution due to the change of the air--water interface 
(projected onto $z=0$):
\begin{eqnarray}
\label{Aproj}
\Delta A_{{\rm proj}, i}
&=&
-\int_0^{2\pi}
d\phi_i\int_{r_{0,\rm ref}(\phi_i)}^{r_0(\phi_i)}
dr\,r \;,
\end{eqnarray}
and a contribution due to the change of the colloid area exposed to fluid I: 
% the %change in the 
%boundary free energy
%%upon height variations of the contact line variation or  tilts of the ellipsoid
%contains a contribution 
%depending on the change
%
\begin{eqnarray}
\label{AI}
  \Delta A_{{\rm I},i}&  =&
 \int_0^{\pi} d\phi' \int_{x_{\rm ref}^{\prime +}}^{{x}^{\prime+}} dx'\!\!
 \int_{y_{\rm ref}^{\prime +}}^{y^{\prime +}} dy'
  \\
 & & \qquad \times
\; \delta
   \left( \phi'- \arctan\frac{y'}{x'} \right) \; \sqrt{g(x',y')}   
\nonumber \\
    & &
  + \int_\pi^{2\pi} d\phi'  \int^{x_{\rm ref}^{\prime -}}_{x^{\prime -}} 
dx' \int^{y_{\rm ref}^{\prime -}}_{y^{\prime -}} dy' 
\nonumber\\&&
\qquad \times \; \delta
   \left( \phi'-\pi- \arctan\frac{y'}{x'} \right) \;\sqrt{g(x',y')} 
\nonumber
\end{eqnarray}
%
% from the change in 
%of the ellipsoid surface area
%exposed to fluid I, and a contribution 
%depending on the change
%%from the change in 
%of the projected meniscus area,
%respectively
%(it turns out, 
%%that it is convenient to transform
%%the surface integral)
In Eq.~(\ref{AI}),
the surface integral
is transformed
 into one over the cartesian components $x',y'$ of a
body--fixed coordinate system with axes
fixed to the main ellipsoid axes 
for computational convenience in taking the derivatives.
The reference contact line is parametrized by $x_{\rm ref}^{\prime +[-]}(\phi')$, 
$y_{\rm ref}^{\prime +[-]}(\phi')$ (for $\phi'<\pi\, [\phi'>\pi]$),
and the shifted contact line is parametrized by $x^{\prime +[-]}(\phi'),
y^{\prime +[-]}(\phi')$.
}

\textcolor{black}{
The calculation of the boundary free energy variation 
$\mc{F}_{{\rm b},i}$  %=\gamma(\cos\theta \Delta A_{{\rm I},i}+\Delta A_{{\rm proj},i})$
can be performed
in two steps:
(1) First, we tilt  ellipsoid $i$ by an angle $\alpha_i$
with a {\it pinned} contact line. Then 
$\Delta A_{{\rm I},i}=0$ holds, but $\Delta A_{{\rm proj},i} \neq 0$.
(2) In a second step, the contact line is released to its
final position $\tilde{v}_{i,\phi_i}$.
In this second step, both $\Delta A_{{\rm I},i}$ and $\Delta A_{{\rm proj},i}$,
contribute to $\mc{F}_{{\rm b},i}$.
The two steps have to be distinguished, since
the orientational tilts change the surface measure of the
ellipsoid. 
In order to avoid the calculation with the
$\alpha_i$-dependent metric 
$g=1+|\nabla z_{i,\rm Ell}(\alpha_i)|^2$
of the ellipsoid surface, we calculate the 
second
variation  $\delta^2 \Delta A_{{\rm I},i}$
 in body--fixed coordinates of the colloid.
The contribution
 $\delta^2\Delta A_{{\rm proj},i}$ from the change
of the projected meniscus area
is determined separately, here the steps
(1) and (2) can be considered together.
In body--fixed coordinates  
the metric is given by
$g'=1+(1-e^2)^2 (x'/z')^2+(y'/z')^2$.
}

\textcolor{black}{
With 
$\tilde{z}_{i,\phi}=\tilde{v}_{i,\phi_i}+u_{\rm ref}(\phi_i)
-h_{i,\rm ref}
%\textcolor{blue}{+\Delta h_i}
$
being the height of the contact line relative to the center of colloid $i$,
the position of the three contact line
in body--fixed coordinates 
after a shift 
$\tilde{v}_{i,\phi_i}$
of the contact line
and a rotation of the ellipsoid by an angle $\alpha_i$
(with respect to the plane  $z=0$)
is determined by the equations 
\begin{eqnarray}
x' &=& \tilde{z} \sin\alpha+x(\phi,\alpha)\cos\alpha \;,
\\
z' &=& \tilde{z} \cos\alpha-x(\phi,\alpha)\sin\alpha \;,
\\
y' &=& y
\end{eqnarray}
parametrising the rotation,
and the constraint
\begin{equation}
F=b^2-{z}^{\prime2}-{y'}^2-(1-e^2)\,{x'}^2\equiv 0
\end{equation}
which ensures that the contact line 
is on the ellipsoid surface.
}

\textcolor{black}{
Employing the relations given above we can 
calculate the contact line position on the rotated ellipsoid and, in
particular, its partial derivatives.
After some algebra, we finally arrive at the
expressions
\bea\label{Ellcoeffzz}
R_{zz}(\phi_i)
&=&
\frac{b^2\sin^2\theta}{\ro(\phi_i)^2(1-e^2\cos^2\phi_i)^2}\;,
\\
\label{Ellcoeffza}
R_{z\alpha} (\phi_i)
&=&
-\frac{1}{1-e^2\cos^2\phi_i}
\bigg[
1-\frac{\zo(\phi_i)^2/\ro(\phi_i)^2}{1-e^2\cos^2\phi_i}
\bigg]
 \nonumber\\&&
 {}
 +\frac{\cos^2\theta\,
\left[(1-e^2)+b^2/r_{0,\rm ref}(\phi_i)^2
 \right]
}{(1-e^2\cos^2\phi_i)^2}
\;,
\eea
and
}

\textcolor{black}{
\bea
\label{Ellcoeffaa}
\lefteqn{R_{\alpha\alpha}(\phi_i)
=}
\\ \nonumber
&&
\frac{1}{1-e^2\cos^2\phi_i}\Bigg[
% \nonumber\\&&
\ro(\phi_i)^2\cos^2\phi_i
%\nonumber\\&&
%\qquad\qquad\qquad
{}
-\frac{\zo^2\,(1+e^2\cos^2\phi_i)}{1-e^2\cos^2\phi_i}
\Bigg]
%\nonumber
\\&&
{}
+  \frac{\cos^2\theta %\,[\cos^2\phi_i+\cos\phi_i\,\sin\phi_i]
}
{2\,(1-e^2\cos^2\phi_i)^3}
\Bigg[
 \nonumber\\&&
-(1-e^2)(b^2-\zo(\phi_i)^2)\,(1+e^2\cos^2\phi_i(1-2e^2\sin^2\phi_i))
 \nonumber\\&&
 {}%\quad\qquad\qquad
+2b^2\left(
1-e^2\cos^2\phi_i
%-e^2\sin^2\phi_i\cos^2\phi_i\,(1-e^2)
\right)\,(1+\cos^2\phi_i\,(1+e^2\sin^2\phi_i))
%-(1-e^2)\,(b^2-z_0(\phi_i)^2)
\Bigg]
\;,\nonumber
\eea
for the coefficients
of the functional Taylor expansion of $\hat{\mc{F}}_{\rm b}$.
Here, $\zo(\phi)^2= b^2 - \ro(\phi)^2[\sin^2\phi +(1-e^2)\cos^2\phi]$.
}

\textcolor{black}{
Note, that  
by minimizing $\hat{\mc{F}}_{\rm b}$ with respect to $\alpha_i$
we ensure
torque balance of the colloid
% with respect to rotations out of the undisturbed interface plane 
%in principal
whereas mimimizing with respect to the height $\Delta h_i$  guarantees that
the total vertical force exerted by the meniscus on the
contact line vanishes.
The numerical calculation of the equilibrium meniscus between
two ellipsoidal particles and the resulting capillary interactions
shows, however, that the effect of $\alpha_i$ is rather small,
leading to changes of the results ${\textstyle
 {\lower 2pt \hbox{$<$} \atop \raise 1pt \hbox{$ \sim$}}} 1\%$
as compared to the computations using the boundary contribution~(\ref{eq:bc}),
in which $\alpha_i$ is neglected.
}

\end{document}